\documentclass{article}
\usepackage{graphics}      % standard graphics specifications
\usepackage{graphicx}      % alternative graphics specifications
\usepackage{longtable}     % helps with long table options
\usepackage{url}           % for on-line citations
\usepackage{bm}            % special 'bold-math' package
\usepackage[usenames]{color}
\usepackage[dvipsnames]{xcolor}
\usepackage{cancel}
\usepackage{soul}
\usepackage{amsmath,amssymb}
\usepackage{epstopdf}
\usepackage{color}
\usepackage[utf8]{inputenc}
\usepackage{epsfig,amsopn,amsmath}
\usepackage{multirow,bigdelim}
\usepackage{subcaption}
\usepackage{graphicx}  % remove 'demo' option for your real document
\begin{document}
	
	\title{Critical Exponents of  
		Master-Node Network Model}
		
	\author{Antonio Mihara\\
			Physics Department,  Federal  University of S\~ao Paulo,
		Diadema--SP, Brazil\\
		Anderson A.\  Ferreira\\
		Physics Department,  Federal  University of S\~ao Paulo,
		Diadema--SP, Brazil\\
		Andr\'e C.\ R.\ Martins\\
		School of  Arts, Science and Humanity,  University of S\~ao Paulo,
		S\~ao Paulo-SP, Brazil\\
		Fernando F. Ferreira\\
			ferfff@usp.br\\
	    Physics Department of FFCLRP, University of S\~ao Paulo,
		Ribeir\~ao Preto-SP, Brazil
	}%

\date{}
\maketitle
	
%%% MIHARA: alteraçõese comentários em VERMELHO	
	
	\begin{abstract}
The dynamics of competing opinions in social network play an important role in society, with many applications in diverse social contexts as consensus, elections, morality and so on. Here we study a model of interacting agents connected in networks to analyze their decision stochastic process.  We consider a first-neighbor interaction between agents in a one-dimensional network with a shape of ring topology. Moreover, some agents are also connected to a hub, or master node, that has preferential choice or bias. Such connections are quenched. As the main results we observed a continuous non-equilibrium phase transition to an absorbing state as a function of control parameters. By using the finite size scaling method, we analyzed the static and dynamic critical exponents to show that this model probably cannot match any universality class already known.
\end{abstract}
	\maketitle

\section{Introduction}

%Opinion dynamics \cite{castellanoetal07,latane81a,galametal82,galammoscovici91,sznajd00,deffuantetal00,martins08a}
Absorbing state phase transitions with quenched disorder have attracted considerable attention \cite{hinrichsen2000non}. Such  phase transitions
can be found in many areas, including physics, chemistry,social systems and biology,
with examples in quantum systems \cite{gutierrez2017experimental}, catalysis \cite{ziff1986kinetic}, surface and interface growth \cite{tang1992pinning},opinion dynamics  \cite{bagnoli2002opinion}, ecology \cite{borile2013effect} and spreading of epidemics \cite{pastor2015epidemic}. Such nonequilibrium phase transitions phenomena has been witnessing considerable theoretical progress, but  many issues remain to be addressed in the literature and surprises may arise.

The critical behavior of absorbing phase
transitions can be categorized into a finite number of universality classes. The most important universality class of absorbing-state transitions is the called directed percolation (DP)\cite{grassberger1979reggeon}, which is a fundamental class of nonequilibrium phase transitions, playing a similar role as the Ising universality class in equilibrium statistical mechanics. Just to give an example of a particular DP model, consider the spreading disease process. 
One may consider a network of (infected or healthy) agents spreading the disease through the neighborhood. Such a system may exhibit an emerging phase transition between endemic or pandemic regime. 

%%%
In the last years fews experimental achievements of the DP class have been observed \cite{hinrichsen2000possible}. However, the main difficulty in verifying the critical exponents of this class is related to inherent impurities present in real experiments. Because such disorders effectively change the intensity of local interactions. And, according to the Harris criterion \cite{noest1986new}, the introduction of a spatial quenching in the DP class is relevant for dimensions $\leq 3$. That is, the spatial correlation length, as well as, the correlation functions scale logarithmic , modifying the values of the critical exponents, as observed by numerical simulation in two dimensions ( see \cite{wada2017critical} and references therein) and confirmed by Hoobergys et al. \cite{hooyberghs2004absorbing} for the contact process model with quenched dilution in one and two dimensions.

 
%%%%
Recently it was introduced a model of dynamics of opinions that assumes a network topology in the form of a ring with some nodes connected to a node outside the ring called a master node that has a fixed opinion on a subject. The connections between the ring nodes and the master node are quenched. Additionally, the dynamics has two absorbing states $(0$ or $1)$, where $1$ is the opinion of the master node and $0$ is the selfish opinion placed as a natural bias of agents in the social network. This model calls attention to presenting a second-order phase transition \cite{ferreira2020stochastic}. %%% COLOCAR A REF. + IMPORTANTE!

In this article, we will explore this model by making a more rigorous analysis of the critical exponents, both static and dynamic.As the main result, we find examples of critical exponents that depend on the model parameters and are different from the exponents of the voter model or directed percolation classes.

This paper is organized as follows. In section II, we describe the model. Then, we present the results of the study based on finite size scaling for static and dynamic critical exponents. Section IV shows the discussion and finally in section V the conclusions.

\section{The Master--Network Model}
\label{sec:model}

\begin{figure}[htb]
		\begin{center}
			\includegraphics[width=0.4\textwidth]{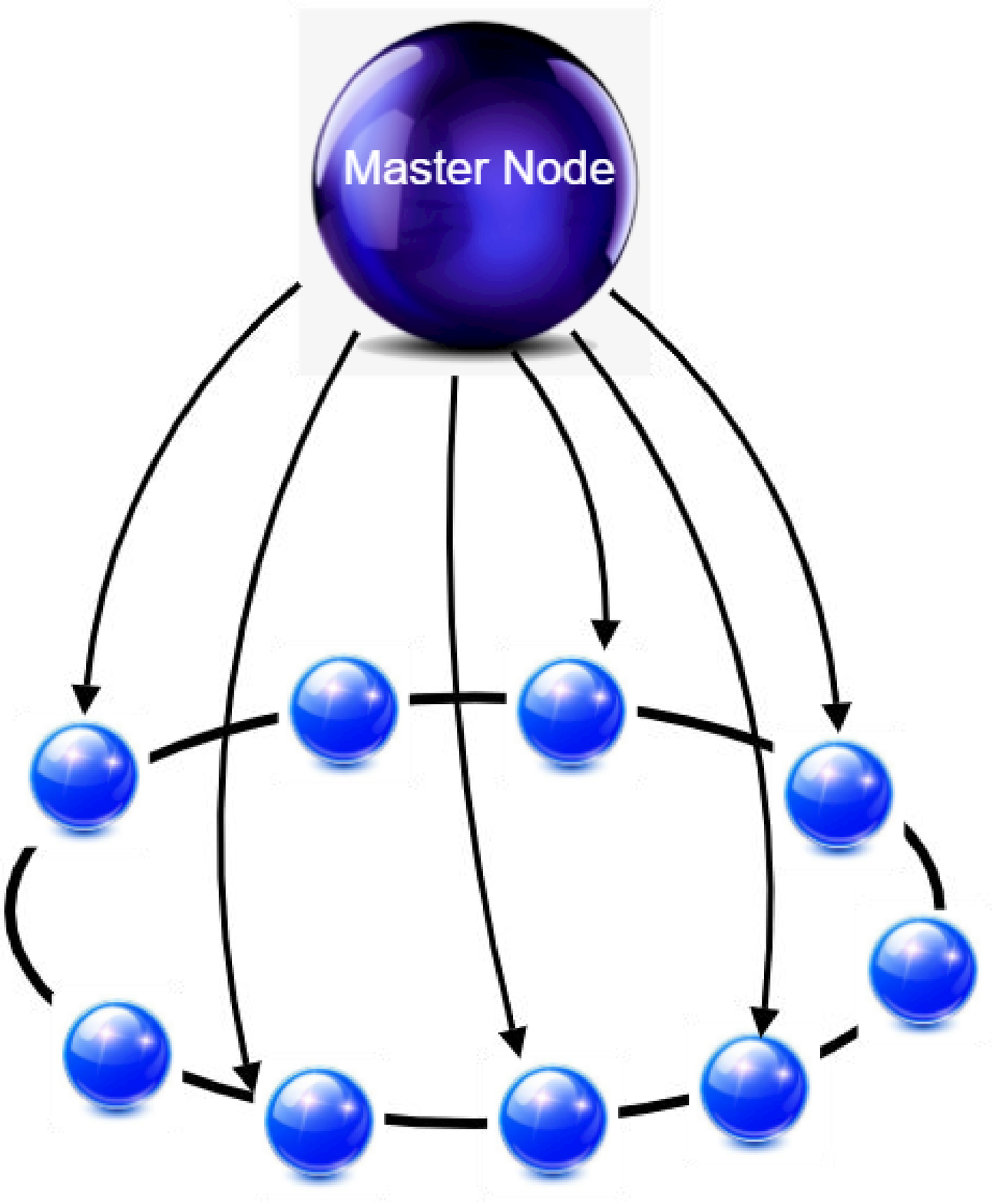}
		\end{center}
		\caption{ Model representation. Small spheres represent interacting nodes (citizens). Each node is in the state $s=0$ or 1. The large sphere represents the master node and its interaction strength with other particles (the citizens) is fixed (denoted by $r$) and the links are in quenched disorder, with density denoted by $\rho$.}
		\label{fig:model}
	\end{figure}

 The model consists of a network in the shape of a ring formed by 
 nodes
 with periodic boundary conditions and a master node as illustrates in  Figure \ref{fig:model}. 
 Initially, each node $i$ represents a particle or agent which may assume two different states $s_i=0,1$ with probability $w_s$. 
The state $s_i=0$  represents, for instance, some opinion or behavior with negative meaning like immoral, selfish, or corrupt behaviors, which harm the collective welfare, while $s_i=1$  represents the opposite.  %So, in principle, this is a particle property. 

%%%%%%%

Each node interacts with its first nearest neighbors and may interact with a master node  with probability $\rho$ (quenched disorder). This connection is drawn at the beginning and is fixed. In other words, $\rho$ is the fraction of the master node's followers. The interaction strength between the master and each connected node is denoted by $r$, which represents the master node's influence power.

The general configuration of the system is given by $(s_i, \Gamma_i)$, with $i=1, \dots, L$ representing the individual nodes and $\Gamma_i = 1 \, (0)$ if there is a connection between it and the master node (or not). In turn, the probability of the node/citizen being influenced by the master node is equal to $ r $.

We are interested here in consensus-seeking agents. That fact limits the choices of possible interaction parameters. However, it is interesting to start considering the general case. 

Assuming the central agent is not influenced by its neighbors, we can write the interaction matrix for all other agents, on the absence of the external field
as summarized in Table \ref{tabcomp}.

\begin{table}[!htb]
	\centering
	\caption{Transition probabilities in the general cases for $\Gamma_i=0$.}
	\begin{tabular}{|cc|c|c|c|c|c}
		\cline{3-6}
		\multicolumn{2}{c|}{}&\multicolumn{4}{|c|}{$s_i \rightarrow s_i^{\prime}$} \\
		\hline
		$s_{i-1}$  & $s_{i+1}$        & $0 \rightarrow 0 $                     & $0 \rightarrow 1 $                 & $1 \rightarrow 0$                 & $1 \rightarrow 1 $    \\
		\hline
		0      &    0         &$ p_1 $              & $1-p_1   $               & $p_2   $               &$1-p_2  $     \\
		\hline
		1      &    1         & $p_3 $               & $1-p_3 $                  & $p_4 $                  & $1-p_4$      \\
		\hline
		0      &    1         & {$1-w_0$} & {$w_0$} &{$w_1$} & {$1-w_1$} \\
			\hline
		1      &   0         & {$1-w_2$} & {$w_2$} &{$w_3$} & {$1-w_3$}                &   \\
		\hline
	\end{tabular}
	\label{tabcomp}
\end{table}

The two last lines of the matrix in Table \ref{tabcomp} correspond to a case where the agent is surrounded by one neighbor with $s_j=0$ and another  with $s_j=1$. In most applications, those two cases are symmetrical and identical, but it is worth noticing that we could introduce a chirality effect.
For this paper, we assume there is no chirality and, therefore, $w_0 = w_2$ and $w_1 = w_3$. That is, the two last lines in the matrix are identical.

It is also interesting to map how traditional models translate in Table \ref{tabcomp} parameters. Normal majority models \cite{galam2004contrarian,de2005spontaneous,martins2010importance,hong2011kuramoto,gambaro2017influence,dong2018survey} where both states are identical would have $p_1=1-p_4=1$ and, since both states are symmetrical, $p_2=1-p_3$. In the simplest case, where majority always convince and it is the only way to change the opinion of an agent, we would have $p_1=p_2=1$. We also have that $w_0=w_1=0$, reflecting the fact there is no update when no majority is observed. In the case all agents were contrarians \cite{galam2004contrarian,de2005spontaneous,martins2010importance,hong2011kuramoto,gambaro2017influence,dong2018survey}, we would have $p_1=p_2 \leq 0.5$, meaning each agent would tend to disagree with the majority. And we can get Galam-like models \cite{galam1986majority,galam1990social,galam2020tipping}, where there is a status quo state (here, $s_i=0$) that tends to be preserved in discussions (with the difference only the central agent can change its opinion), if we make $p_1=p_4=0$, $p_2=1-p_3$, $w_0=0$, and $p_2 > w_1\neq 0$. Traditionally, we would also have $p_2=1$ in this case, but that is not necessary. Another interesting scenario corresponds to $w_1=1-w_0$, $p_1=p_2$, and $p_3=p_4$. In this situation, the central agent has no memory of its previous state, as its new state depends only on the neighbors and not on its initial preference. 

The introduction of a central agent connected to some agents allows us to have agents influenced by their neighbors using distinct rules. That difference can, in principle, be used to describe many different scenarios. In the original paper and in much of the analysis that follows, this framework was used to explore a scenario where a central agent tries to impose the choice $s=1$. While the terminology in the previous paper \cite{ferreira2020stochastic} identifies that option as a moral rule, the same model can also correspond to other scenarios, such as compliance to law, obedience to a norm, or even the adoption of a specific product.
	
We suppose that the interaction between node $i$ and 
its neighbors $i-1$ and $i+1$ depends on the value of $\Gamma_i$. For $\Gamma_i=0$, $s_i$ will align with the majority in the neighborhood, a situation which leads to consensus. If there are differences in opinions between neighbors (frustration), the state $s_i = 0\, (1)\rightarrow s_i^{\prime}=1\, (0) $ with probability $w_0 \, (w_1)$ as we can see in Table \ref{tab1}.

\begin{table}[!htb]
	\centering
	\caption{Transition probabilities for $\Gamma_i=0$.}
	\begin{tabular}{|cc|c|c|c|c|c}
		\cline{3-6}
		\multicolumn{2}{c|}{}&\multicolumn{4}{|c|}{$s_i \rightarrow s_i^{\prime}$} \\
		\hline
		$s_{i-1}$  & $s_{i+1}$        & $0 \rightarrow 0 $                     & $0 \rightarrow 1 $                 & $1 \rightarrow 0$                 & $1 \rightarrow 1 $    \\
		\hline
		0      &    0         &$ p_1 $              & $1-p_1   $               & $p_2   $               &$1-p_2  $     \\
		\hline
		1      &    1         & $p_3 $               & $1-p_3 $                  & $p_4 $                  & $1-p_4$      \\
		\hline
		0      &    1         & \multirow{2}{*}{$1-w_0$} & \multirow{2}{*}{$w_0$} & \multirow{2}{*}{$w_1$} & \multirow{2}{*}{$1-w_1$}\ \\
		1      &   0          &                   &                   &                   &   \\
		\hline
	\end{tabular}
	\label{tab1}
\end{table}

For the case $\Gamma_i=1$, the transition probabilities 
will be different. If both of its first nearest neighbors are $s_{i-1}  =
s_{i+1}= 0$ then the particle will change from state $s_i=0 \, (1)$ to $s_i^{\prime}=1$,
with probability $ q \, (r) $.  If both of its first nearest neighbors are $s_{i-1}  = s_{i+1}= 1$, then the particle will change from state $s_i=0 \, (1)$ to $s_i^{\prime}=1$ with probability $1$. Finally, if there is frustration between neighbors, the state $s_i = 0\, (1)\rightarrow s_i^{\prime}=1\, (0) $ with probability $r_0 \, (r_1)$, respectively. This case is summarized in Table \ref{tab2}. 
% 
% We summarize all the situations described above in tables 
% \ref{tab1} and \ref{tab2}, where we present the
% probabilities (for each case $\Gamma_i=0,1$) that the node $i$
% will change its state $s_i \rightarrow s_i^{\prime}$ according to the state of its first nearest neighbors.

\begin{table}[h]
	\centering
	\caption{Transition probabilities for $\Gamma_i=1$.}
	\begin{tabular}{|cc|c|c|c|c|c}
		\cline{3-6}
		\multicolumn{2}{c|}{}&\multicolumn{4}{|c|}{$s_i \rightarrow s_i^{\prime}$} \\
		\hline
		$s_{i-1}$  & $s_{i+1}$        & $0 \rightarrow 0 $                     & $0 \rightarrow 1 $                 & $1 \rightarrow 0$                 & $1 \rightarrow 1 $    \\
		\hline
		0      &    0         & $1-q_1$               & $q_1$                  & $1-r$                  & $r$       \\
		\hline
		1      &    1         & $q_2 $               & $1-q_2 $                  & $q_3 $                  &$ 1-q_3$      \\
		\hline
		0      &    1         & \multirow{2}{*}{$1-r_0$} & \multirow{2}{*}{$r_0$} & \multirow{2}{*}{$r_1$} & \multirow{2}{*}{$1-r_1$} \\
		1      &   0          &                   &                   &                   &   \\
		\hline
	\end{tabular}
	\label{tab2}
\end{table}

It is worth mentioning that, in principle, the central agent can also work as a marker for the heterogeneity of the agents. Heterogeneity, of course, does not need this framework, but it is nice to notice that it can be included here. For example, if we have normal interactions when  $\Gamma_i=0$, we can easily introduce inflexibles  \cite{galam2005heterogeneous,martins2013building,crokidakis2015inflexibility,mobilia2015nonlinear,galam2020tipping} if we make, when  $\Gamma_i=1$, $q_1=q_3=r_0=r_1=0$ and $q_2=r=1$. 

	Aiming to analyse this model, we choose some constraints to the parameters $r_0, r_1, w_0$ and $w_1$ in terms of $r$, $\rho$  and $\Delta=w_0-w_1$. 
	
	The parameter $\Delta$ measures the intrinsic bias (or tendency) of an agent be in the state $s=0 (1)$ when $\Delta>0 (<0)$ in the absence of any interaction or influence.
	If $\Delta$ is positive, the system has a tendency to be opposite to the master node. This is the most interesting situation. So, we 
	study the regime in which $0<\Delta<1$.
	
	%%%%%%%%%%%%%%%%%%%%
	
	The probabilities $r_0$ and $r_1$ are parameterized so that when the master's influence is null ($r=0$), we have $r_0=w_0$ and $r_1=w_1$. Otherwise, when $r=1$, we should have necessarily $r_0=1$ and $r_1=0$. The simplest way is through a linear relation:
	\begin{equation}
	r_0=r+(1-r)(\frac{1-\Delta}{2})  \, ,
	\label{eq:r0}
	\end{equation}
	\begin{equation}
	r_1=(1-r)(\frac{1+\Delta}{2})  \, .
	\end{equation}
	
Finally, the parameterized version of the model has only three free parameters:
	\begin{equation}
	0\leq r \leq 1,;\;\;\;\;0\leq \Delta \leq 1,;\;\;\;\mbox{and}\;\;\;\;\; 0\leq\rho\leq 1.
	\end{equation}
	
\section{Results}

The most important characteristic of the critical phenomena is the set of critical exponents related with the singular  behaviour   of the thermodynamics functions near the critical point \cite{stanley1999scaling}. In equilibrium statistical mechanics the principal scheme of classification of the continuous phase transitions in two 
dimesions started to be structured after the seminal work 
of ref.~\cite{belavin1984infinite}. 
However in non-equilibrium phase transitions the situation is more complex, 
 due to the extra dimension time and other degrees of 
freedom present in dynamics, other critical exponents (dynamical exponents) are necessary to classify a non-equilibrium phase transition \cite{marro2005nonequilibrium,odor2004universality, lubeck2004universal, hinrichsen2000non}

\subsection{Static Exponents}
\noindent

Monte Carlo simulations were used for obtaining the critical 
exponents for the order parameter $I$, the average density of immoral citizens. By setting the values of $\Delta$ and $\rho$, one can expect
a power--law behavior $I(r) \sim | r - r_c |^{\beta} $, near the critical
point $r_c$. For large $L$ and $| r - r_c |\rightarrow 0$,  
$I$ can be written in the following form:
\begin{equation}
I \sim L^{-\beta/\nu_{\bot}}\, f\left( (r-r_c).L^{1/\nu_{\bot}} \right) \, ,
\end{equation}{}
with the scaling function $f(x) \sim |x|^{\beta}$, for large $|x|$.

\begin{figure}

  \includegraphics[width=10cm,height=8cm,keepaspectratio]{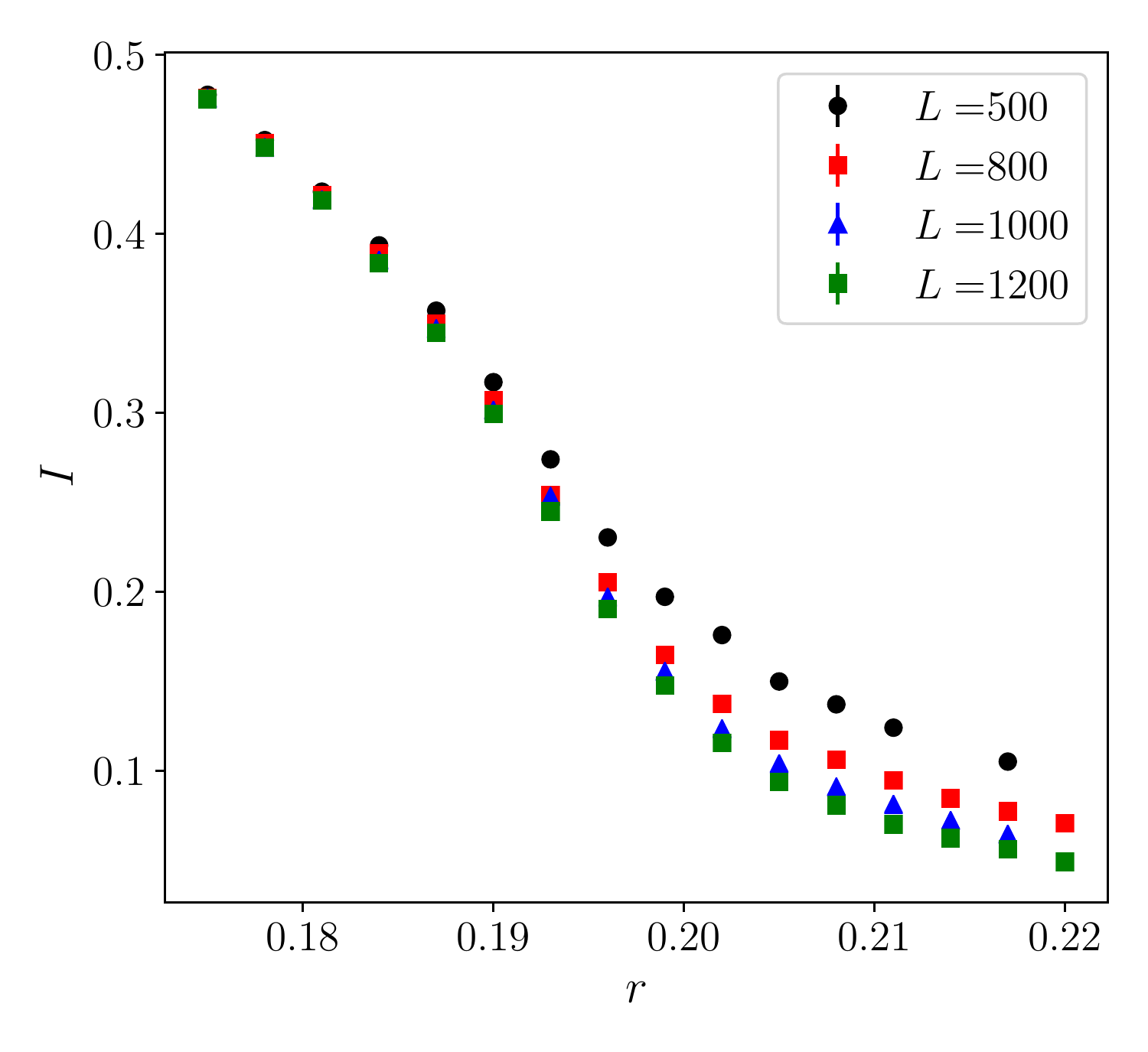}\\\includegraphics[width=10cm,height=8cm,keepaspectratio]{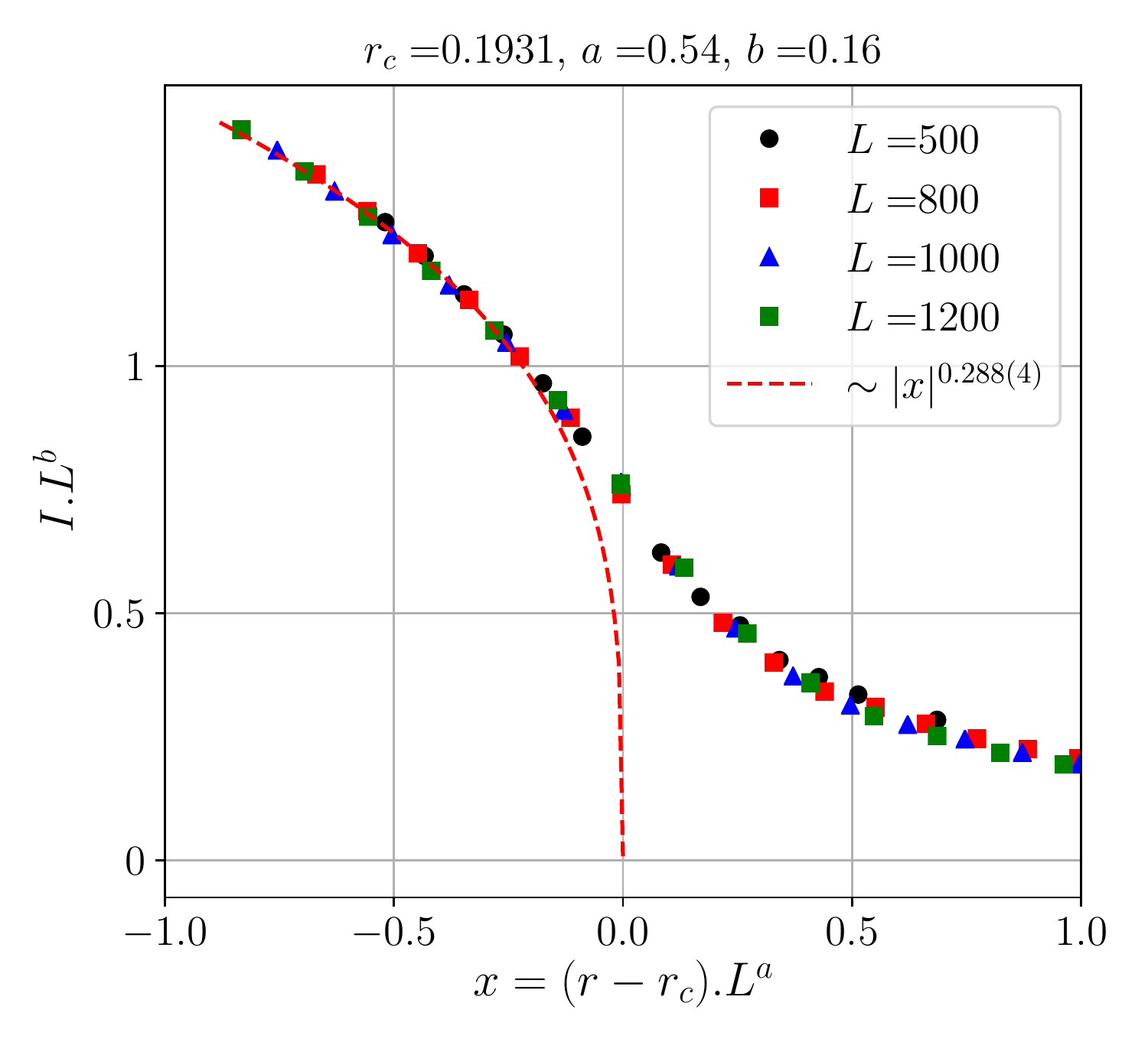}
  \caption{Top: $I\, \times\,r$, for parameters $\Delta=0.5, 
    \,\rho = 0.7$, for different network sizes.Bottom: finite-size scaling. The dashed red line shows the behavior}
    \label{fig:2}
\end{figure}

We performed numerical simulations for $\Delta=0.5$ with 
two different values of $\rho$ ($\rho=0.7$ and $\rho=0.9$) and obtained estimates for $\beta$ and $\nu_{\bot}$ for both cases. In Fig.\ref{fig:2}
we present (on the top panel) the results obtained with $\Delta=0.5$
and $\rho=0.7$ for different network sizes; 
and the corresponding finite--size scaling (FSS) analysis on 
the right panel.

%\begin{figure}[b] %[hbt]
    %\centering
%    \includegraphics[width=0.5\textwidth]{figuras/Fig2A.pdf}\\
%     \includegraphics[width=0.5\textwidth]{figuras/Fig2B.pdf}
 %   \caption{Left: $I\, \times\,r$, for parameters $\Delta=0.5, 
 %   \,\rho = 0.7$, for different network sizes. Right: finite-size scaling. The dashed red line shows the behavior 
 %   $f(x) \sim |x|^{0.288(4)}$, for large $|x|$.}
 %   \label{fig:SE}
%\end{figure}{} 
%\twocolumngrid

%%%%%%%%%%%%%%%%

%%%%%%%%%%%%%%%%%%%%%%%%%%%%%%%

From FSS we obtained the critical point $r_c = 0.1931(9)$, and the
exponents $a \equiv 1/\nu_{\bot} = 0.54(9)$ and 
$b \equiv \beta/\nu_{\bot} = 0.160(12)$. From both results 
one can obtain the following estimates: $\beta = 0.30(5)$ and
$\nu_{\bot}=1.9(3)$.

However on the bottom panel of Fig.\ref{fig:2} we observe that the 
scaling function $f$ behaves as $f(x) \sim |x|^{0.288(4)}$, for large $|x|$ (red dashed line), then we can obtain a more precise estimate: $\beta = 0.288(4)$.

For the parameters $\Delta=0.5$ and $\rho=0.9$, we obtained
$r_c=0.1426(9)$,  $a \equiv 1/\nu_{\bot} = 0.530(8)$ and 
$b \equiv \beta/\nu_{\bot} = 0.1393(18)$. From these exponents 
we obtained the following estimates: $\beta = 0.263(5)$ and
$\nu_{\bot}=1.887(28)$.

\subsection{Dynamic Exponents}
\noindent

Simulations were performed with initial configuration very near the
absorbing state: all sites with $s_i=1$, except eleven
sites in the ``center'' of the network (the central site,
$i=L/2$, and five sites both sides) with $s_i=0$.
Three time--dependent quantities were measured at the critical point: 

\begin{itemize}
    \item $N(t)=$ the mean number of immoral citizens (sites with 
    $s=$0).
    \item $P(t)=$ the mean survival probability of immoral citizens.
    \item $R^2(t) = $ the mean quadratic distance of immoral citizens from
    the center ($i=L/2$) of network.
\end{itemize}{}

For the asymptotic evolution we observed power--law behavior: $N \sim t^{\eta}$, $P(t)\sim t^{-\delta}$ and $R^2(t) \sim t^z$. In Fig.~\ref{fig:DE} we present our results for
$\Delta=0.5, \rho = 0.7$ and some values of $r$.
We obtained the exponents: $\eta = 0.3272(9)$, 
$\delta = 0.0684(9)$ and $z = 0.930(3)$, for $r = 0.1923$.

%\vspace{0.5cm}
%%%%%%%%%%%%%%%%%%%%%%%%%%%%%%%%%%%%%%
%% MODIFICADO EM 02/09/2020
%% RESULTADOS RECENTES:

We also studied the case with $\Delta=0.5, \rho = 0.9$.
Now for $r=0.1434$, the exponents obtained: $\eta = 0.4228(16)$, 
$\delta = 0.0798(7)$ and $z = 1.164(7)$.

Finally one can expect the following 
 {\it scaling/hyperscaling} relations \cite{marro2005nonequilibrium}
\begin{equation}
    \frac{\beta}{\nu_{\bot}} = 2\,\frac{\delta}{z} \, , \quad 
    2\eta + 4\delta = d z \, .
\end{equation}{}
In table \ref{tab:results} we summarize the results above, and in table \ref{tab:scaling} we observe that both scaling relation above are satisfied, within the error estimates.

\begin{table}[ht]
\caption{Static and dynamic exponents obtained for $\Delta=0.5$, for different values of $\rho$.}
\label{tab:results}
\centering
\begin{tabular}{c||cc|ccc} %|SS}
$\rho$ & $\beta$ & $\nu_{\bot}$ & $\eta$ & $\delta$ & $z$  \\ % & $\beta/\nu_{\bot}$ & $2\delta/z$  \\
\hline \hline
0.7 & 0.30(5)  & 1.9(3)    & 0.3272(9)  & 0.0684(9) & 0.930(3) \\ % & 0.160(12)  & 0.1684(7) \\
\hline
0.9 & 0.263(5) & 1.887(28) & 0.4228(16) & 0.0798(7) & 1.164(7)    % & 0.1393(18) & 0.1371(10) 
\end{tabular}

\end{table}

\begin{table}[ht]
\caption{Scaling relations for the exponents.}
\label{tab:scaling}
\centering
\begin{tabular}{c||ccc|ccc}
$\rho$ & $\beta/\nu_{\bot}$ &=& $2\delta/z$ &  $2\eta+4\delta$  &=& $z$ \\
\hline\hline
0.7   & 0.160(12)           & &  0.1471(20)  &  0.928(4)    & &  0.930(3) \\
\hline
0.9   & 0.1393(18)          & &  0.1371(14) &  1.165(4)    & &  1.164(7)
\end{tabular}
\end{table}

%\section{ Power-Law X Power-Log behavior}

%\begin{figure}[hbt]
 %   \centering
 %   \includegraphics[width=0.49\textwidth]{CurvasNxt_rho0p7.pdf}
%    \includegraphics[width=0.49\textwidth]{CurvasNxlnt_rho0p7.pdf}
%    \caption{$N$ vs $t$ (left) and $N$ vs $\ln t$ (right) plots 
%    for $10 < t \leq 10^5$, both in log-log scale for different values of $r \sim r_c$. }
%    \label{fig:plots}
%\end{figure}

%\newpage

%\section{Comparison with the exponents of 2dXY$h_4$ model}

\section{Discussions}

\begin{figure}[hbt]
    \centering
    \includegraphics[width=0.50\textwidth]{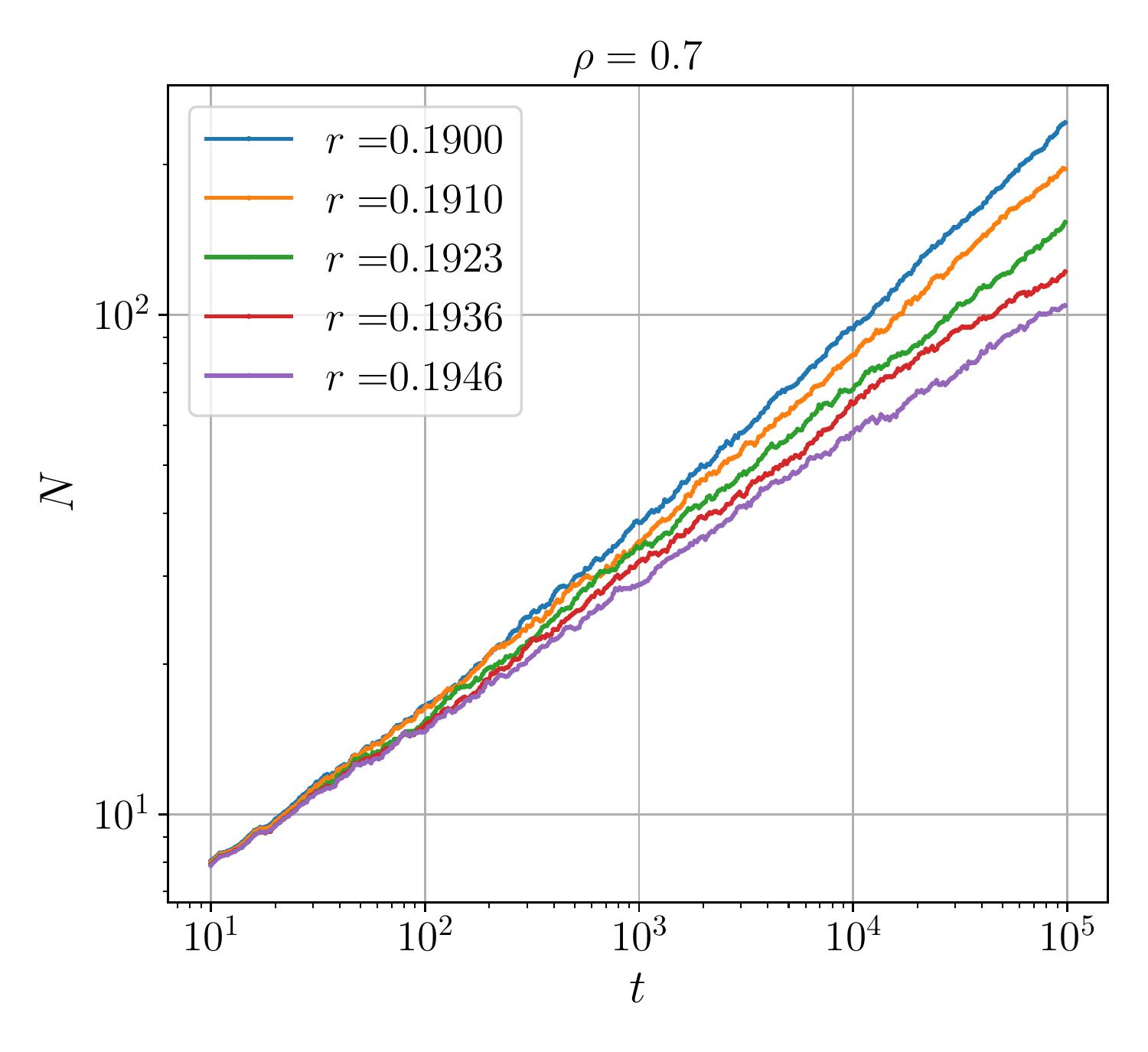}
    \caption{$N(t)$ for $\Delta=0.5$, $\rho=0.7$ and some values of $r$.}
    \label{fig:DE}
\end{figure}

An important issue in obtaining dynamic exponents concerns the time behavior of the quantities $N, P, R^2$: at the critical point, such quantities depend on $ t $ as a power law or as a power of $\ln t$? In Figure \ref{fig:DE} one can observe that $N$ behaves as a power law of $t$, since the critical curve (corresponding to $r\approx r_c$) is very close to straight line on a log-log plot.

Quenched models are not always relevant compared to pure (non-quenched) models, i.e, they do not bring fundamental novelties already present in pure models. According to Harris \cite{noest1986new} a quenched perturbation is relevant when the relation $\nu_{\perp}d<2$ is satisfied. In this model, the non-quenched versions ($\rho=0$ or $\rho=1$) have no transition. Therefore, we have a relevant quenched model that does not satisfy the Harris criterion.

One aspect to be questioned is whether the critical exponents of the model under study belong to some already known universality class. In a first analysis we could try identifying these critical exponents with some class already well known in the literature of one-dimensional stochastic processes with absorbing states, such as the DP class \cite{grassberger1979reggeon}, the parity conserving (PC) universality class \cite{PhysRevLett.68.3060}, the universality class of pair contact process with diffusion \cite{Howard_1997},\cite{Odor_2003}, the Manna universality class in sandpile models \cite{Manna_1991} and  the voter universality class \cite{PhysRevLett.87.045701} since those models are the promising ca

For comparison , we will highlight the exponent set for the case where the density of connected followers is $\rho=0.9$. At this choice, the static critical exponents are $\beta=0.263(3)$ and $\nu_{\perp}=1.887(2)$, while in the DP class at $d=1$ we have $\beta=0.276486( 8)$ and $\nu_{\perp}=1.096854(4)$. Likewise, there is a discrepancy in the dynamics critical exponents.

Another possibility would be to compare these results with the quenched DP university class. However, as shown analytically by Hoobergys et al. \cite{hooyberghs2003strong} through the quenched renormalization group, an introduction of an impurity (quenched perturbation) in dimension $d=1$ in the DP class establishes that the exponent $\nu_{\perp}=2$ and the exponent $\beta=\frac{3-\sqrt{5}}{2}$. Furthermore, as argued by these same authors, the temporal correlation functions scale logarithmically instead of a power law, as observed in our numerical analysis for our model in these parameterizations.

%%%%%%%%%%%%%%%%%%%%%%%%%%%%%%%%%%%%%%

\begin{figure}[hbt]
    \centering
    \includegraphics[width=0.49\textwidth]{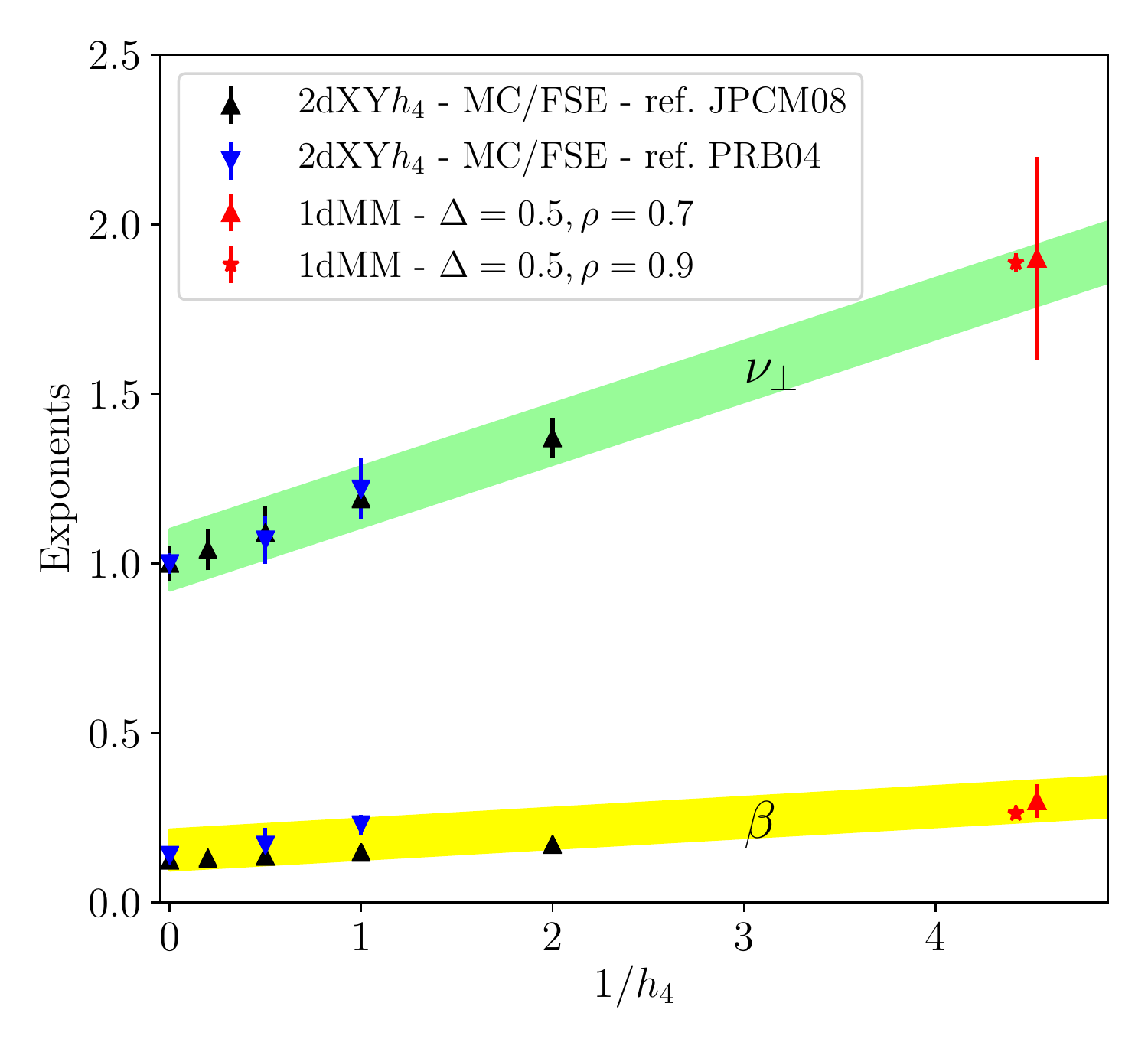}
    \caption{Comparison between our exponents (red points) and those
    of the $2dXYh_4$ model (blue and black points).}
    \label{fig:expXYh4}
\end{figure}

Alternatively, a priori, one can map a nonequilibrium stochastic system in $d$-dimensions to a quantum Hamiltonian \cite{schutz1995reaction} which in turn can be related to a $(d+1)-$dimensional equilibrium system \cite{hertz1976prb}. So it seems plausible that such a coincidence is not accidental. We could not prove the connection between our (1$d$) model and the 2$dXYh_4$ model, but we speculate that such a connection probably exists and we put the exponents ($\beta, \nu$) and those of the 2$dXYh_4$ model \cite{taroni2008universal} on the same plot (Fig.\ref{fig:expXYh4}

As it was predicted in ref.\cite{jose1977renormalization} 
and presented in Fig.5 of ref.\cite{taroni2008universal}, the exponents of
2$dXYh_4$ model depends linearly on $1/h_4$. In Fig.\ref{fig:expXYh4} 
we plot the exponents $\beta, \nu$ of 2$dXYh_4$ model corresponding to 
to different values of $1/h_4$ (blue and black triangles), such results were obtained through several Monte Carlo/finite size scaling studies, see ref.\cite{taroni2008universal} and references therein.The yellow and green bands are to guide the eye and to reinforce that $\beta$ and $\nu$ are proportional to $1/h_4$.
By assuming that a mapping between both models is possible, we also plot the results (red symbols) of our 1$d$ model corresponding to some hypothetical values of $h_4$.

%%%%%%%%%%%%%%%%%%%%%%%%%%%%%%%%%%%%%%%

%{\color{red} 

\section{Conclusions}

The model described and studied here introduces a central agent that can influence a proportion of agents and, in that sense, it is similar to other models where the effects of media and well-connected agents \cite{mckeownsheehy06a}, \cite{martins2008central} or norms \cite{Elsenbroich2014}  were studied. However, the central agent here can not only influence the opinions of those connected to it. It can also change how connected agents are influenced by their neighbors. In that sense, it can act as a catalyst. And, as we have seen, we can describe numerous distinct models of the literature using the framework presented here. Therefore, the model can also be seen as an effort to generalize and compare the existing literature \cite{martins12b,Boettcher2017,Galam2020}.

But the model is not interesting only for opinion dynamics. Given its dynamic properties, it seems to be novel and relevant for the non equilibrium statistical mechanics literature, as far as we know.

To determine whether the model is something new in relation to many others that are already well studied, we attempted to identify an eventual universality class in which it could be classified. Surprisingly, it doesn't map onto existing classes. Furthermore, critical exponents have a dependency on model parameters. This feature makes the analysis quite laborious to be conducted numerically. The ideal approach is to seek analytical approaches that allow a broader understanding of the model.

Finally, it is worth remembering that the solutions presented were obtained for a very particular choice of constraints between the parameters described in the  tables \ref{tab1} and \ref{tab2} . This parameterization was just to simplify the analysis. Certainly, other cases may also have even more intriguing properties.

 

%}

\end{document}